\documentclass[12pt,a4paper]{article}
\usepackage{graphicx}
\usepackage{color}
\usepackage{amsmath}

\usepackage{cite}

\newcommand{\EQ}{Eq.~}

\newcommand{\FIG}{Fig.~}
\newcommand{\FIGS}{Figs.~}
\newcommand{\SEC}{Sec.~}

\title{Numerical analysis of a reinforcement learning model with the dynamic aspiration level in the iterated Prisoner's Dilemma}
\bigskip
\author{Naoki Masuda${}^{1,2*}$ and Mitsuhiro Nakamura${}^1$\\
\ \\
\ \\
${}^{1}$ 
Department of Mathematical Informatics,\\
The University of Tokyo,\\
7-3-1 Hongo, Bunkyo, Tokyo 113-8656, Japan
\ \\
${}^2$
PRESTO, Japan Science and Technology Agency,\\
4-1-8 Honcho, Kawaguchi, Saitama 332-0012, Japan\\
\ \\
$^*$ Corresponding author (masuda@mist.i.u-tokyo.ac.jp)}

\begin{document}
\setlength{\baselineskip}{24pt}

\maketitle

\newpage

\begin{abstract} \setlength{\baselineskip}{24pt} 
Humans and other animals can adapt their social behavior in response to environmental cues including the feedback obtained through experience. Nevertheless, the effects of the experience-based learning of players in evolution and maintenance of cooperation in social dilemma games remain relatively unclear. Some previous literature showed that mutual cooperation of learning players is difficult or requires a sophisticated learning model. In the context of the iterated Prisoner's Dilemma, we numerically examine the performance of a reinforcement learning model. Our model modifies those of Karandikar \textit{et al.} (1998), Posch \textit{et al.} (1999), and Macy and Flache (2002) in which players satisfice if the obtained payoff is larger than a dynamic threshold. We show that players obeying the modified learning mutually cooperate with high probability if the dynamics of threshold is not too fast and the association between the reinforcement signal and the action in the next round is sufficiently strong. The learning players also perform efficiently against the reactive strategy. In evolutionary dynamics, they can invade a population of players adopting simpler but competitive strategies. Our version of the reinforcement learning model does not complicate the previous model and is sufficiently simple yet flexible. It may serve to explore the relationships between learning and evolution in social dilemma situations.
\end{abstract}

Keywords: cooperation, direct reciprocity, Prisoner's Dilemma, reinforcement learning

\newpage

\section{Introduction}\label{sec:introduction}

Human beings and other animals often cooperate with each other even in
social dilemma situations where to not cooperate is apparently
a rational choice.  A standard framework in which social
dilemma situations are studied is the Prisoner's Dilemma game (PD) and its
variants. Many theoretical mechanisms for emergence and
maintenance of cooperation in social dilemma games have been reported thus far
\cite{Axelrod1984book,Boyd1985book,Nowak2006book,Sigmund2010book}.

Most of these mechanisms do not deal with the adaptation or learning
of individuals. We use the term learning to refer to individual
learning (i.e., experience-based adaptation),
but not to social learning (i.e., imitation).
Learning implies that an individual takes advantage
of the history of the games that it has played
to perform better in subsequent rounds.
A learning individual changes behavior on the basis of some
statistics of the game results.
Laboratory experiments suggest that humans do learn
during sequences of games \cite{Camerer2003book,Glimcher2009book}.
The learning of the social behavior of animals, including humans, has been modeled
in various game and non-game situations
\cite{Rapoport1965book,Cross1983book,Boyd1985book,Fudenberg1998book,Camerer2003book} 

Learning in a game is relevant only in an
iterated game.
It is well known that mutual cooperation can be optimal
in the iterated PD \cite{Trivers1971,Axelrod1984book}.
Action rules that have mainly been considered in the context of
the iterated PD are those that do not adjust conditional probabilities of cooperation
upon experience. A player
using a look-up table
that relates the next action to the 
outcome of the game in the current and past few rounds belongs to this class
\cite{Axelrod1984book,Kraines1989,Nowak1989AMC,Nowak1990Acta,Nowak1990TPB,Lindgren1991ALife,Nowak1992Nature_gtft,Nowak1993Nature,Nowak2006book,Sigmund2010book}. This important class
includes well-known strategies such as the tit-for-tat (TFT).
However, the flexibility
of such a strategy appears to be limited.

Players using reinforcement learning, on which we focus in this study,
exploit information about past encounters to adapt
the probability of cooperation conditioned by the outcome of the game in a couple of past rounds.
Because of their flexibility,
such learning players may be strong competitors in the
iterated PD.  If learning players compete relatively well in a
population, the learning behavior may spread to become dominant in the
population through evolutionary dynamics.  Nevertheless, the possible
roles of reinforcement learning in the iterated PD, either in favor of or against
the promotion of cooperation, are relatively unexplored.  In fact,
players using reinforcement learning have generally been unsuccessful in the PD and other
social dilemma games
\cite{Macy1996SMR,Sandholm1996BioS,Posch1999RoyalB,Taiji1999PhysicaD,Macy2002pnas,MasudaOhtsuki2009BMB}.
Although an artificial neural network model, for example, enables
mutual cooperation \cite{Gutnisky2004AL}, such a complicated mechanism
may not be implemented by humans or other animals.  It seems that
the current understanding of social dilemmas is mostly based on studies in the fields of
evolutionary biology and economics.  Because experience-based learning, 
and reinforcement learning in particular,
is quite evident in humans and other animals, both in terms of
behavior and neural activities \cite{Camerer2003book,Glimcher2009book}, 
clarifying
the role of reinforcement learning in the iterated PD may provide an additional
understanding of how subjects cope with social dilemmas.

In the present study, we numerically
examine a variant of the reinforcement learning model \cite{Karandikar1998JET,Posch1999RoyalB,Macy2002pnas}
in the iterated PD. Following Macy and Flache (2002), we call the original
model the Bush--Mosteller (BM) model.
A player obeying the BM reinforcement learning
(BM player for short) would continue an action (i.e., cooperate or
defect) after gaining
a relatively large payoff and would switch the
action otherwise.
If the threshold payoff above which the player satisfices, which is
called the aspiration level, is fixed,
BM players can mutually cooperate
\cite{Rapoport1965book,Macy1991AJS,Macy1996SMR,Posch1999RoyalB,Macy2002pnas,Izquierdo2007GEB,Izquierdo2008JASSS}. 
The BM player with the fixed aspiration level studied in these articles
is essentially the same as
Pavlov that only uses the information about the
immediate past \cite{Kraines1989,Nowak1993Nature}.
Pavlov is known to be exploited by the unconditional
defector and behave too generously to the unconditional cooperator.

Real subjects may adapt the aspiration level in response to the
results of the game \cite{Simon1959AER}.  The BM model with the
adaptive aspiration level is not known to yield a large probability of
mutual cooperation except in some limited cases
\cite{Karandikar1998JET,Posch1999RoyalB,Macy2002pnas}.  We remark that
performance of other reinforcement learning models with the adaptive
aspiration level have also been investigated in the PD and other games
\cite{Pazgal1997IJGT,Kim1999EconTh,Palomino1999IJGT,Dixon2000JEBO,Borgers2000IER,Oechssler2002JEBO,Bendor2003APSR,Napel2003GEB,Cho2005JET}.
In the temporal difference learning, which is a dominant form of
reinforcement learning in the brain, dopamine neurons represent the
difference between the obtained reward and the dynamic expected reward that
changes according to the subject's experience
\cite{Schultz1997Science,Montague2002Neuron,Daw2006CONB,Glimcher2009book}.
The reinforcement signal in the BM model with the adaptive aspiration level
is given by the difference between the obtained reward and the dynamically changing
aspiration level such that the BM model with the adaptive aspiration level is at least loosely connected
to neural evidence.

We show that a simple modification of the BM model with the adaptive aspiration level
drastically
changes the behavior of the player. The modified BM player mutually
cooperates with a large probability and is competitive in
evolutionary dynamics. The modification is done such that
the reinforcement signal is reflected to the action selection in the
next round fairly strongly. The aspiration level must adapt with
a low to intermediate learning rate for sustaining cooperation. It should be noted that
our modification to the BM model does not 
introduce an additional complexity to the original
BM model with the adaptive aspiration level \cite{Karandikar1998JET,Posch1999RoyalB,Macy2002pnas}.

\section{Model}

We consider the symmetric two-person PD whose payoff matrix is
given by
\begin{equation}
\bordermatrix{
 & {\rm C} & {\rm D} \cr
{\rm C} & R & S \cr
{\rm D} & T & P \cr}, \;
\label{eq:payoff}
\end{equation}
where $T>R>P>S$ and $R>(T+S)/2$.  The entries of \EQ\eqref{eq:payoff}
represent the payoffs that the row player gains. Each row (column)
corresponds to the action of the row (column) player, i.e.,
cooperation (C) or defection (D).  Because $T>R$ and $P>S$, mutual
defection is the only Nash equilibrium of the single-shot game.
Unless otherwise stated, we assume a standard payoff matrix for the PD
given by $R=3$, $T=5$, $S=0$, and $P=1$.

A pair of players play the PD for a predetermined number of
rounds denoted by $t_{\max}$.  We denote the round by $t$ ($=1$, 2,
$\ldots$).  Although the Nash equilibrium of the iterated PD is
defection in all the rounds, which can be derived by backward induction,
we assume for simplicity that players do not carry out backward
induction. We could avoid this technical subtlety by assuming that a
next round occurs with a certain probability such that the last round
is not known beforehand \cite{Axelrod1984book,Nowak2006book}.

To model a learning player, we use a variant of the BM
reinforcement learning model adapted to the game situation,
pioneered in Rapoport and Chammah (1965).
Our model is a variant of the BM model with
the adaptive aspiration level \cite{Karandikar1998JET,Posch1999RoyalB,Macy2002pnas}.

In round $t$, the BM player intends to cooperate with probability $p_t$.
We set the initial condition to $p_1 = 0.5$.
In addition, we assume that the player misimplements the action (i.e.,
C or D) to play the opposite action with a small probability
$\epsilon$.
The payoff that the BM player gains in round
$t$ is denoted as $r_t\in \{R,T,S,P\}$.
We define the stimulus, or the reinforcement signal,
using the sigmoidal function as
\begin{equation}
s_t = \tanh\left[\beta(r_t - A_t)\right],
\label{eq:stim}
\end{equation}
where $A_t$ is the aspiration level in round $t$ above which the BM
player satisfices. The degree of
satisfaction is parametrized by $s_t$, and $-1<s_t<1$ holds true.
If $s_t>0$ ($s_t<0$), the BM player is motivated to
keep (switch) the current action in the next
round.  The sensitivity of the stimulus to the reinforcement signal
$r_t-A_t$ is parametrized by $\beta\ge 0$.

The dynamics of the probability of cooperation are given by
\begin{equation}
p_{t+1} = \begin{cases}
p_t + (1-p_t)s_t, & (\mbox{Action in round } t= {\rm C}, \mbox{ and } s_t\ge 0),\\
p_t + p_t s_t, & (\mbox{Action in round } t= {\rm C}, \mbox{ and } s_t < 0),\\
p_t - p_ts_t, & (\mbox{Action in round } t= {\rm D}, \mbox{ and } s_t\ge 0),\\
p_t - (1-p_t) s_t, & (\mbox{Action in round } t= {\rm D}, \mbox{ and } s_t < 0).
\end{cases}
\label{eq:p_t update}
\end{equation}
Finally, the dynamics of the aspiration level are given
by
\begin{equation}
A_{t+1}=(1-h)A_t + h r_t,
\label{eq:dynamics of A}
\end{equation}
where $h$ represents the learning rate of the aspiration level, which is also called habituation \cite{Macy2002pnas}.
In contrast to previous models in which $h$ decays as $t$ increases
\cite{Erev1998AER,Cho2005JET}, we assume that $h$ is a fixed constant.
Unless otherwise stated, we set the initial value of $A_t$ to $A_1=(R+T+S+P)/4$, which
is equal to the expected payoff when there are an equal number of
cooperators and defectors in a population. As a remark,
the possibility of cooperation in the iterated PD and other games
was examined when the update of $A_t$ is driven by
the average payoff over time \cite{Kim1999EconTh,Cho2005JET},
the maximal experienced payoffs \cite{Pazgal1997IJGT}, or
the payoff averaged over the population \cite{Palomino1999IJGT,Oechssler2002JEBO}.

The difference between our model and the 
Macy--Flache model \cite{Macy2002pnas} lies in 
\EQ\eqref{eq:stim}. Macy and Flache use
$s_t= \ell(r_t - A_t)/\max[T-A_t,A_t-S]$ instead of \EQ\eqref{eq:stim}.
As described below, this difference results in a remarkable difference in 
the behavior of the player.
In other words, we show that
reacting strongly to the play in the previous round (i.e., large $\beta$)
is necessary for mutual cooperation.
A deterministic decision maker with the 
adaptive aspiration level used in
Posch \textit{et al.} (1999)
corresponds to
$\beta=\infty$.
We numerically show that $\beta$ does not have to be extremely large for mutual cooperation.
We remark that,
if $\beta=\infty$ and
the aspiration level is fixed (i.e., $h=0$),
the strategy is a win-stay lose-shift one.
In particular, our BM model with $\beta=\infty$
and $h=0$ is
equivalent to the Pavlov strategy 
\cite{Kraines1989,Nowak1993Nature} if $P<A_t<R$.

\section{Results}

\subsection{BM versus BM}\label{sec:BM-BM}

In this section, we examine the performance of a BM player
playing against another BM player.
We assume that the two players employ the same values of $\beta$ and 
$h$. For a range of $\beta$ and $h$, the fraction of the rounds in
which the focal BM player cooperates
is shown for three values of implementation error,
$\epsilon=$ 0, 0.01, 0.1, and two values of the number of rounds,
$t_{\max}=100$, 1000, in \FIG\ref{fig:BM-BM}.
The presented values are averages 
over 100 trials in this and the following figures
unless otherwise stated.
The fraction of cooperation is large when $h$ is small and $\beta$ is large.
The results are fairly robust, despite some degradation, even under
10\% of the error in the action implementation (\FIG\ref{fig:BM-BM}(c, f)).
Remarkably, a large fraction of cooperation can be established only after 
$t_{\max}=100$ rounds (\FIG\ref{fig:BM-BM}(d, e, f)).
These results are in contrast to those for other
reinforcement learning models for social dilemma games, where
the establishment of mutual cooperation requires a large number of rounds \cite{MasudaOhtsuki2009BMB}
or is simply difficult \cite{Macy1996SMR,Sandholm1996BioS,Posch1999RoyalB,Taiji1999PhysicaD,Macy2002pnas,MasudaOhtsuki2009BMB}.

In \FIG\ref{fig:BM-BM}, $\beta$ must be larger than approximately $2.7$ 
for the fraction of cooperation to be large for small $h$.
When $\beta$ is in this range,
\EQ\eqref{eq:stim} suggests that the reinforcement signal
$s_t$ would be typically close to $-1$ or $1$ before a possible equilibrium
is reached.
This is because $|r_t - A_t|$ is typically about unity or larger
when $R=3$, $T=5$, $S=0$, and $P=1$. Then,
\EQ\eqref{eq:p_t update} implies that $p_t$ is close to 0 or 1, and
the selection of the action tends to be almost deterministic. This
deterministic nature of the BM player seems to pave the way to
mutual cooperation. This result is consistent with those obtained from
other models of reinforcement learning with the adaptive aspiration level
\cite{Palomino1999IJGT,Oechssler2002JEBO}.

\begin{figure}
\begin{center}
\includegraphics[height=6cm]{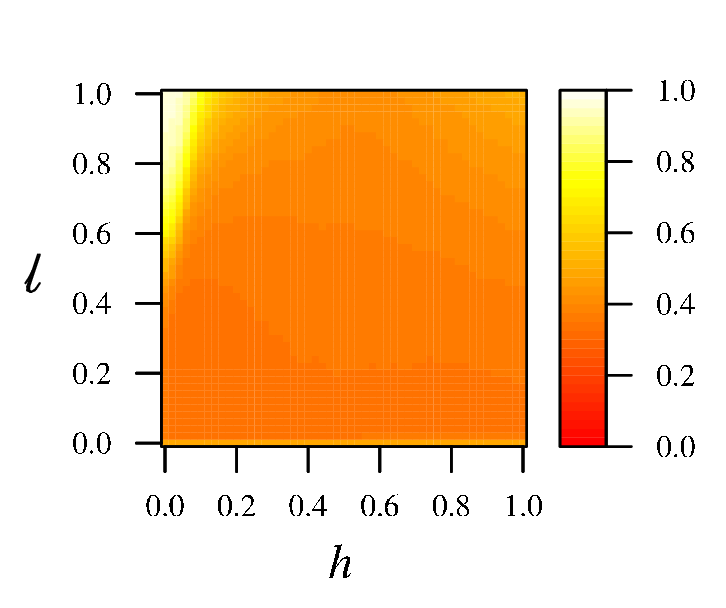}
\caption{\setlength{\baselineskip}{0.77cm}
Fraction of cooperation of the BM player playing against another BM
player in the Macy--Flache model. We set $t_{\max}=1000$, $\epsilon=0$,
$R=3$, $T=5$, $S=0$, and $P=1$, and vary $h$ and $\ell$.}
\label{fig:BM-BM Macy}
\end{center}
\end{figure}

Some mutual cooperation also occurs in the Macy--Flache original BM
model with the adaptive aspiration level \cite{Macy2002pnas}.
For the sake of comparison, the fraction of cooperation 
in the Macy--Flache model with $\epsilon=0$ and $t_{\max}=1000$
is shown in \FIG\ref{fig:BM-BM Macy} for
various values of $h$ and the sensitivity to the stimulus
$\ell$. The fraction of cooperation is much smaller
than that for our model. We consider that 
this is because the stimulus $s_t$ with which
to update the probability to cooperate in the next round is not sufficiently
sensitive to the reinforcement signal $r_t-A_t$ in the Macy--Flache model. To satisfy
$-1\le s_t\le 1$ such that \EQ\eqref{eq:p_t update} is well-defined,
we need $\ell\le 1$.
Then, $s_t$ would not be close to $-1$ or $1$ in
a considerable number of rounds. Then, the action in the next round
is not likely to be very sensitive to the result of the game in the current round.
Regardless of the value of
$\ell$, Macy's model roughly corresponds to
our model with a small value of $\beta$. This interpretation is
consistent with the result that a small $\beta$
yields a small fraction of cooperation in our model (\FIG\ref{fig:BM-BM}).

Our results are also consistent with those in
Posch \textit{et al.} (1999),
in which the authors analytically showed that mutual cooperation is
difficult when $h=1$ and $\beta=\infty$ (their YESTERDAY strategy)
and that mutual cooperation is established if the temptation payoff $T$
is not too large when $h$ is tiny and $\beta=\infty$ (their FARAWAY
strategy). A sufficiently small $h$ combined with slight stochasticity 
in the dynamics of $h$ also leads to mutual cooperation \cite{Karandikar1998JET}.
Our numerical results extend their analytical results
in showing that the BM players mutually cooperate
up to an intermediate value of $h$ if $\beta$ is sufficiently large.

\begin{figure}
\begin{center}
\includegraphics[height=6cm]{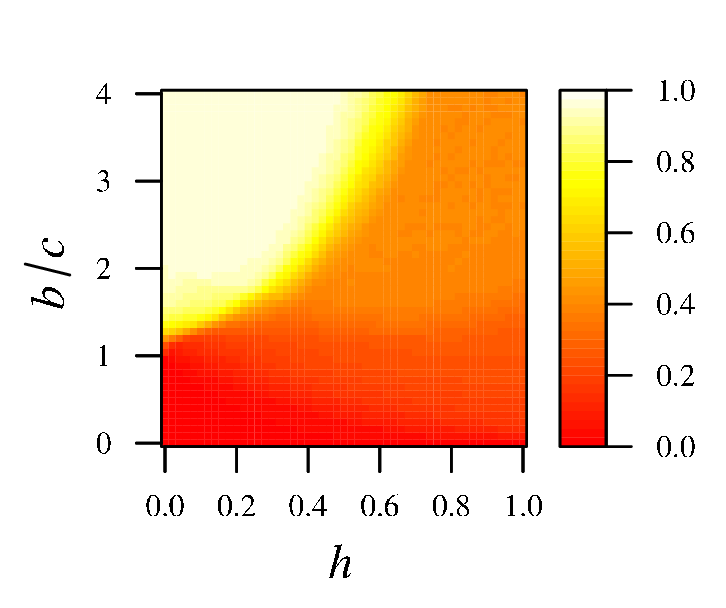}
\caption{\setlength{\baselineskip}{0.77cm}
Fraction of cooperation of the BM player playing against another BM
player. We set $t_{\max}=1000$, $\epsilon=0.02$, $\beta=3$, $c=1$,
$R=b-c$, $T=b$, $S=-c$, and $P=0$, and vary $h$ and $b/c$.}
\label{fig:BM-BM b/c}
\end{center}
\end{figure}

To test the robustness of the results against changes in the payoff
matrix, we set $R=b-c$, $T=b$, $S=-c$, and $P=0$, and measure the fraction
of cooperation as a function of $h$ and the benefit-to-cost ratio $b/c$.
We set $\beta=3$, for which the BM player mutually cooperates
when $R=3$, $T=5$, $S=0$, $P=1$, and $h$ is small (\FIG\ref{fig:BM-BM}).  The
results for $t_{\max}=1000$, $\epsilon=0.02$, and $c=1$ are shown in
\FIG\ref{fig:BM-BM b/c}.  The cooperation decreases with an increase
in $h$. Nevertheless, the threshold value of $b/c$ above which the BM
players mutually cooperate with a probability close to unity
differs only slightly up to $h\approx 0.25$. 

\begin{figure}
\begin{center}
\includegraphics[height=6cm]{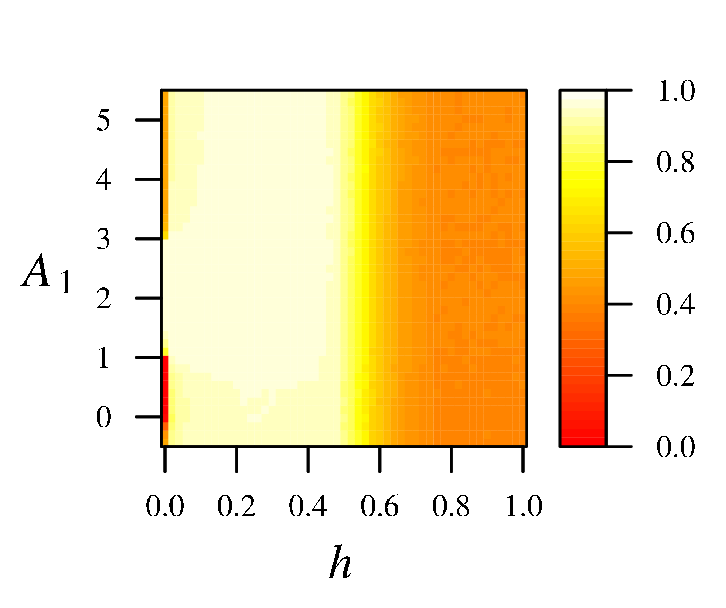}
\caption{\setlength{\baselineskip}{0.77cm}
Fraction of cooperation of the BM player playing against another BM
player. We set $t_{\max}=1000$, $\epsilon=0.02$, $\beta=3$,
$R=3$, $T=5$, $S=0$, and $P=1$, and vary $h$ and $A_1$.}
\label{fig:BM-BM vary h and A_0}
\end{center}
\end{figure}

A small $h$ requires
a relatively large number of rounds before the cooperative
equilibrium is reached, even if the parameter values are set to yield a
cooperative equilibrium. The fraction of cooperation when 
$t_{\max}=1000$, $\epsilon=0.02$, $\beta=3$, $R=3$,
$T=5$, $S=0$, and $P=1$ is shown in
\FIG\ref{fig:BM-BM vary h and A_0} for various values of $h$ and 
initial aspiration level $A_1$.
Figure~\ref{fig:BM-BM vary h and A_0} suggests that
$h> 0.03$ is necessary for $h$ to relax to an equilibrium value
within $t_{\max}=1000$ rounds. When $h$ is too small,
the fraction of cooperation strongly
depends on $A_1$. If $P<A_1<R$,
the BM player is essentially the same as Pavlov for such a small $h$.
In this case, mutual cooperation is realized, reflecting the fact that
Pavlov cooperates against itself \cite{Kraines1989,Nowak1993Nature}.
However, if we start from
a different $A_1$, the fraction of cooperation would be small
for a small value of $h$.

\begin{figure}
\begin{center}
\includegraphics[height=6cm]{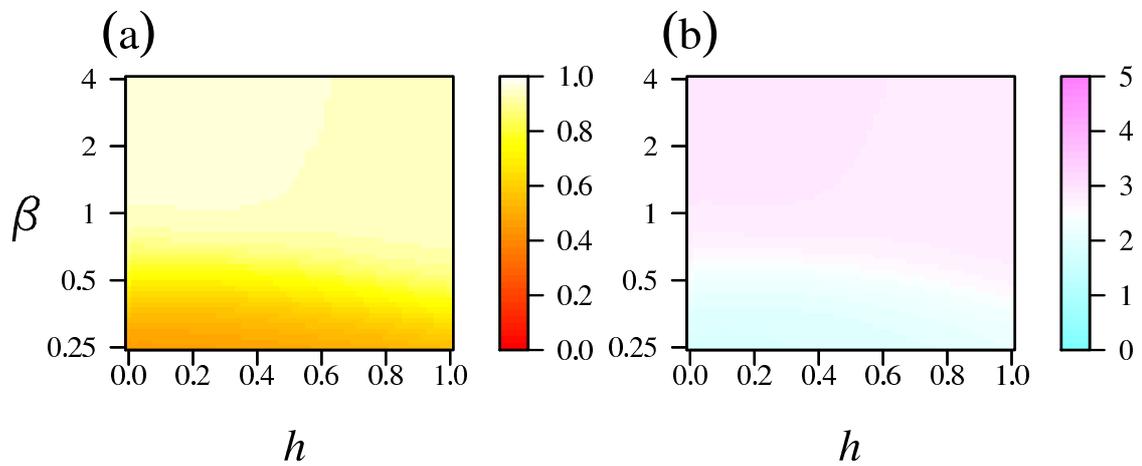}
\caption{\setlength{\baselineskip}{0.77cm}
Behavior of a focal BM player with $h=0.3$ and $\beta=3$ against a BM opponent
with different values of $h$ and $\beta$.
(a) Fraction of cooperation and (b) mean payoff of the focal BM player.
We set $t_{\max}=1000$, $\epsilon=0.02$, $R=3$, $S=0$, $T=5$, and $P=1$.}
\label{fig:BM hetero}
\end{center}
\end{figure}

Two BM players may have different parameter values.  Because
  \FIG\ref{fig:BM-BM vary h and A_0} suggests that the value of $A_1$
  is irrelevant to the fraction of cooperation unless $h$ is too
  small, we set $A_1=(R+T+S+P)/4$ and examine the case in which two
  players have different values of $h$ and $\beta$.  We set $h=0.3$
  and $\beta=3$ for a focal BM player. For the opponent BM player with
  the identical values of $h$ and $\beta$, \FIGS\ref{fig:BM-BM} and
  \ref{fig:BM-BM b/c} guarantee that the two players mutually cooperate.
  With $t_{\max}=1000$ and $\epsilon=0.02$, the fraction of
  cooperation and the mean payoff for the focal player when the opponent has different
  values of $h$ and $\beta$ are shown in
  \FIG\ref{fig:BM hetero}(a) and \ref{fig:BM hetero}(b), respectively.
  Figure~\ref{fig:BM hetero}(a) indicates that the focal BM player
  mostly cooperates with the opponent with similar values of $h$ and
  $\beta$.  Although the fraction of cooperation is small when the
  opponent has small $\beta$, the focal BM
  player avoids being exploited by the opponent in this way 
(\FIG\ref{fig:BM hetero}(b)).  In both cases, the
  focal BM player performs well against the BM opponent.

\subsection{BM against reactive strategies}\label{sub:reactive}

We examine the behavior of the BM player against
players adopting the reactive strategy.
A reactive
strategy is an often used non-learning strategy, and it is
specified by two parameters $p$ and $q$ ($p, q\in [0,1]$)
and the initial condition.
The reactive player cooperates with probabilities $p$ and $q$
when the opponent cooperates and defects in the previous
round, respectively \cite{Nowak1989AMC,Nowak1992Nature_gtft,Nowak2006book}.
 Unconditional cooperation (ALLC), unconditional defection
(ALLD), and TFT correspond to $(p,q)=(1,1)$, $(0,0)$,
and $(1,0)$, respectively. We assume that a player with the reactive strategy 
cooperates in the first round.

\begin{figure}
\begin{center}
\includegraphics[height=12cm]{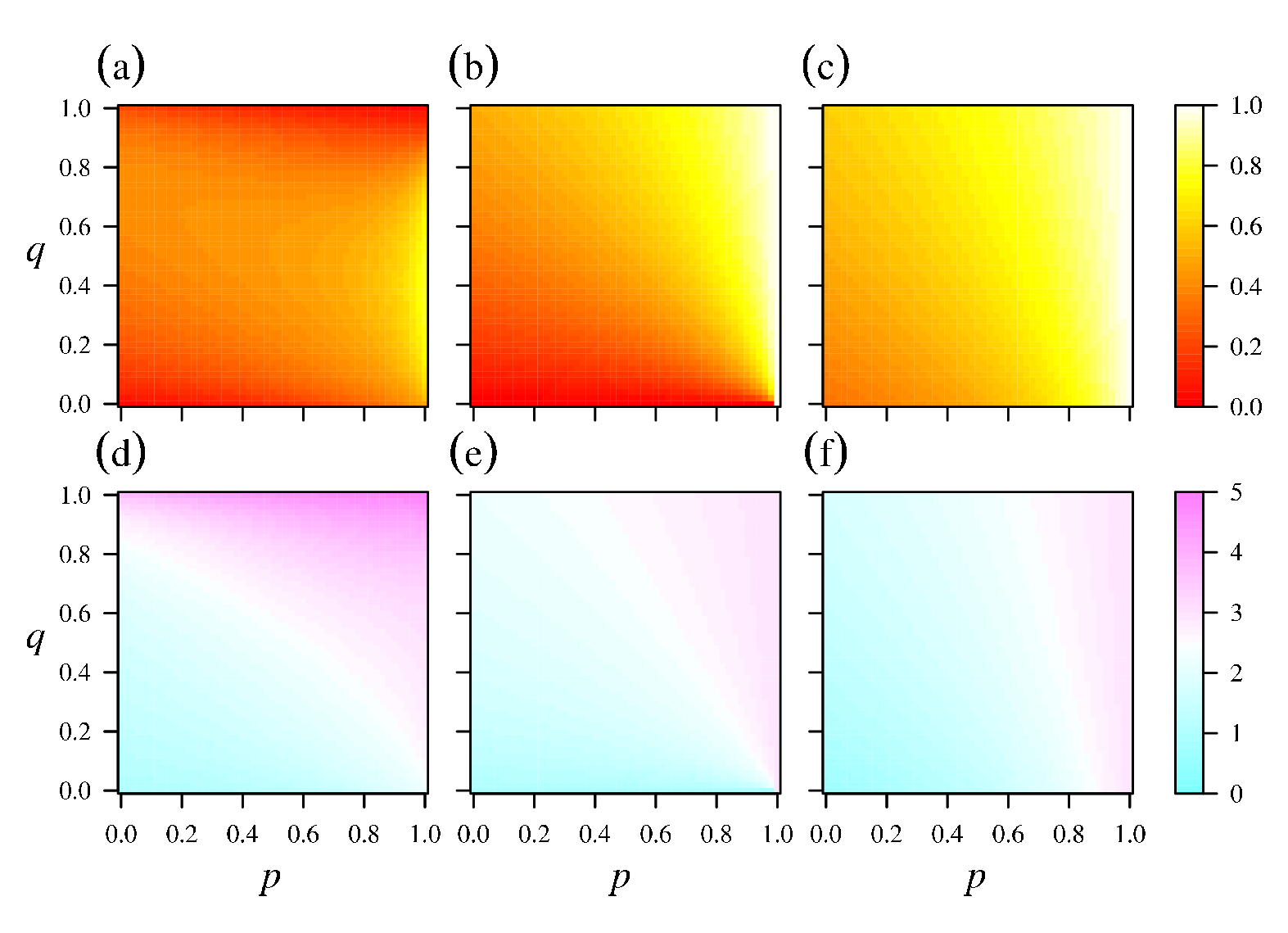}
\caption{\setlength{\baselineskip}{0.77cm}
Fraction of cooperation 
of (a) BM player, (b) TFT, and (c) GTFT against reactive strategies.
We set $t_{\max}=1000$, $\epsilon=0.02$, $h=0.3$, $\beta=3$,
$R=3$, $S=0$, $T=5$, and $P=1$. Mean payoff of (d) BM player,
(e) TFT, and (f) GTFT against reactive strategies.}
\label{fig:BM-reactive}
\end{center}
\end{figure}

The fraction of cooperation of the BM player against various 
reactive strategies is shown in
\FIG\ref{fig:BM-reactive}(a) for $t_{\max}=1000$, $\epsilon=0.02$,
$h=0.3$, and $\beta=3$.
The BM player rarely cooperates with ALLC and ALLD.
To never cooperate is the optimal action against these two strategies.
The BM player cooperates with TFT 
in approximately half the rounds. This is not an optimal behavior; perpetual cooperation is optimal
when the opponent is TFT \cite{Axelrod1984book}.
The mean payoff for the BM
player against the reactive strategy is shown in
\FIG\ref{fig:BM-reactive}(d).
For all the values of $p$ and $q$,
the mean aspiration level of the BM player
is indistinguishable from the mean payoff shown in \FIG\ref{fig:BM-reactive}(d).
As already implied in \FIG\ref{fig:BM-reactive}(a),
the BM player 
exploits ALLC and gains more than $4.5$ per round. The BM
is not exploited by ALLD and gains approximately $P=1$ per round.
The BM player gains approximately 2.5 per round against TFT. This value is
smaller than but not too far from 
$R=3$ per round, which would be obtained by mutual cooperation with
TFT.
Figure~\ref{fig:BM-reactive}(a) shows that
the BM player cooperates with a large probability with
generous tit-for-tat (GTFT) defined by $p=1$ and $q=1/3$ for the current payoff matrix.
GTFT is a strong competitor in the iterated PD
\cite{Nowak1992Nature_gtft}. 
Although the BM player occasionally defects against GTFT,
the BM player gains $\approx R=3$ per round, which 
would be obtained by mutual cooperation. 

The BM player does not play optimally against TFT. However,
the BM player is generally strong against reactive strategies,
as compared to TFT and GTFT. To support this, we plot
the fraction of cooperation and the mean payoff for TFT against
the reactive strategy in \FIG\ref{fig:BM-reactive}(b)
and \ref{fig:BM-reactive}(e), respectively. The plotted values are analytical
solutions obtained by Nowak and Sigmund
\cite{Nowak1989AMC,Nowak1990Acta,Nowak1990TPB,Nowak2006book}, which are summarized in
Appendix for completeness. As shown in \FIG\ref{fig:BM-reactive}(b),
TFT does not cooperate with itself because TFT
is intolerant to haphazard defection of the opponent
\cite{Nowak1992Nature_gtft}.
In addition, TFT does not exploit ALLC. This is
why TFT is eventually invaded by ALLD in evolutionary
simulations in which ALLC, ALLD, and TFT coexist
\cite{Nowak1992Nature_gtft,Nowak1993Nature}. The payoff for TFT
against the reactive strategy (\FIG\ref{fig:BM-reactive}(e)) 
is smaller than that
for the BM player (\FIG\ref{fig:BM-reactive}(d)) for a wide range
of $p$ and $q$. This is particularly true
for large values of $p$, which encompass TFT, GTFT, and ALLC.

The fraction of cooperation and the mean payoff for GTFT player
against the reactive strategy are shown in
\FIG\ref{fig:BM-reactive}(c) and \ref{fig:BM-reactive}(f),
respectively (see Appendix for derivation).
The GTFT player performs better than the BM player and TFT for large $p$.
However, GTFT is too generous to
ALLC and ALLD. When
$p$ and $q$ are both small or both large,
the payoff for GTFT (\FIG\ref{fig:BM-reactive}(f)) is smaller than
that for the BM player (\FIG\ref{fig:BM-reactive}(d)). In addition, 
the payoff for GTFT and that for the BM player are comparable
when $q>p$ and when $(p,q)$ is close
to that of GTFT. On the basis of these numerical results, we conclude
that the performance of the BM player against the reactive strategy is
comparable to that of GTFT.

\subsection{Evolutionary simulations}

If the BM player is a strong competitor in the iterated PD,
it should be able to 
evolve in a population in which different strategies
coexist. To examine this point,
we simulate evolutionary dynamics of populations where BM players
and non-learners coexist in the beginning.
We model non-learners by the 
stochastic memory-one strategy with which the player determines an
action based on its own action and that of the opponent in the
previous round \cite{Nowak1995JMB}.

There are four types of outcomes of the pairwise interaction in a round, i.e.,
CC, CD, DC, and DD.
The first and second letters (i.e., C or D) represent the actions of the focal
player obeying the memory-one strategy (memory-one player for short)
and the opponent, respectively.
The memory-one player is parametrized by the action in the
first round and four 
probabilities $p_{\rm CC}$, $p_{\rm CD}$, $p_{\rm DC}$, and
$p_{\rm DD}$. The probability corresponding to the outcome of the present round is 
used as the probability that the memory-one player cooperates
in the next round.
For example, if both players cooperate, the memory-one player
cooperates with probability $p_{\rm CC}$ in the next round.
Initially, the memory-one player
is assumed to cooperate with probability $p_{\rm CC}$.
The memory-one strategy includes many 
important strategies such as the reactive strategy and Pavlov 
\cite{Nowak1995JMB}.
We assume $p_{\rm CC}$, $p_{\rm CD}$, $p_{\rm DC}$, $p_{\rm DD}$ $\in
\{0, 1/m, 2/m, \ldots, (m-1)/m, 1\}$
and that
there are initially 
an equal number of memory-one players of each type.
The case $m=3$ with a slight modification is employed in 
a previous study \cite{Hauert2002jtb_learn}.
To be realistic, we assume that both the memory-one
player and the BM player
misimplement the intended action
with probability $\epsilon=0.02$. 

We denote the number of players in the population by $N$.
Each player in the population plays against each of the
other $N-1$ players iteratively for $t_{\max}=1000$
rounds in a single generation.
The results shown in the following
are qualitatively the same if $t_{\max}$ is reduced to 500.
We normalize the payoff of each player by dividing it by $(N-1)t_{\max}$
such that the payoff per generation falls
between $S$ and $T$.
We update the strategy of the players during evolutionary dynamics
according to the Fermi rule
\cite{SzaboToke1998pre,TraulsenNowakPacheco2006pre}.
At the end of each generation,
we pick a pair of players $i$ and $j$ from
the population with equal probability. We denote their single-generation
payoffs as $r^{(i)}$ and $r^{(j)}$. With probability
$1/\left[1+\exp\left(\tilde{\beta} (r^{(i)}-r^{(j)})\right) \right]$, player $i$
copies the strategy of player $j$. With the remaining probability,
player $j$ copies the strategy of player $i$.
We set $\tilde{\beta}=1$.
If the parent (i.e., player whose strategy is copied)
is the BM player, the child (i.e., player copying the
strategy of the parent) becomes
the BM learner. In this case, both the parent and the child start with
$p_{t=1}=0.5$ and $A_{t=1}=(R+T+S+P)/4$ in the next generation.
If the parent is 
a memory-one player, the child
inherits the parent's parameter values $p_{\rm CC}$, 
$p_{\rm CD}$, $p_{\rm DC}$, and $p_{\rm DD}$.
For simplicity, we do not consider mutations.

\begin{figure}
\begin{center}
\includegraphics[height=6cm]{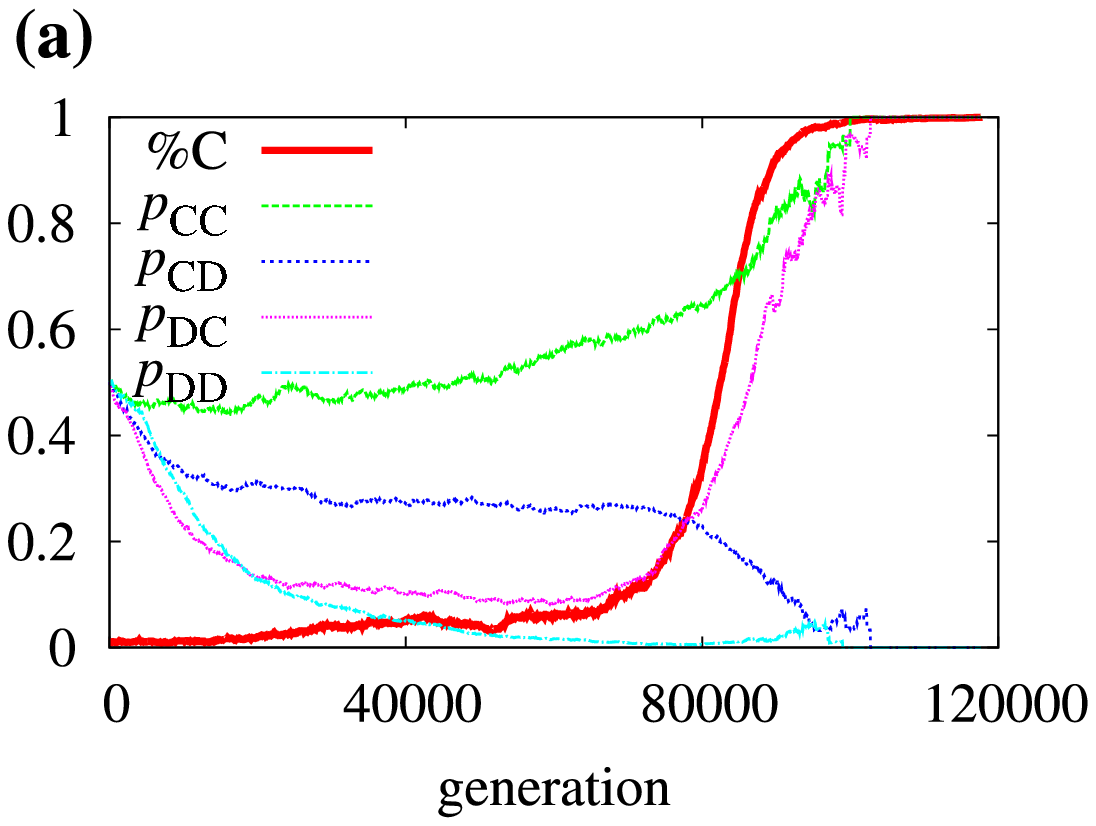}
\includegraphics[height=6cm]{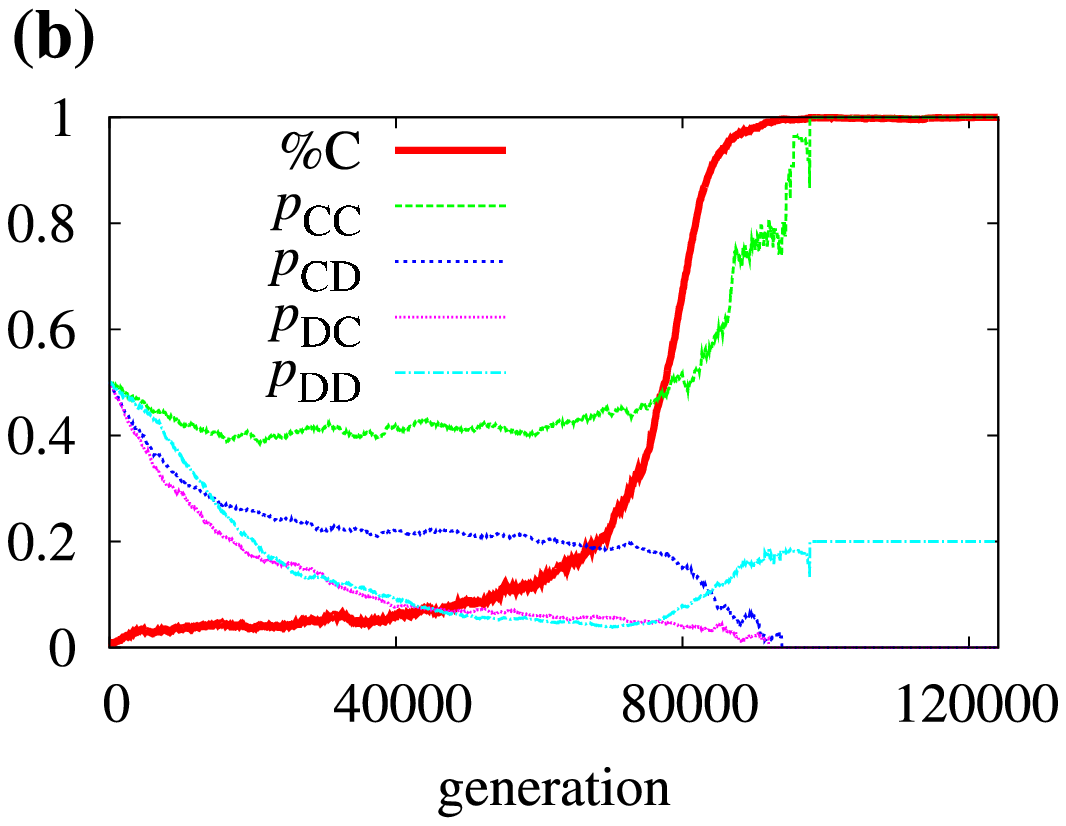}
\caption{\setlength{\baselineskip}{0.77cm}
Time courses of evolutionary dynamics. 
We set $t_{\max}=1000$, $\epsilon=0.02$, $h=0.3$, $\beta=3$,
$R=3$, $T=5$, $S=0$, and $P=1$. 
(a) Results for $N=2585$, $m=3$, and $4^4$ types of memory-one
strategies. (b) Results for $N=2617$, $m=5$, and $6^4$ types of
memory-one strategies.}
\label{fig:evolutionary}
\end{center}
\end{figure}

To examine the possibility
that the BM player invades a population of players with various
memory-one strategies,
we start evolutionary simulations with 1\% of the BM players in a population.
Two time courses of typical runs when $h=0.3$ and $\beta=3$
are shown in \FIG\ref{fig:evolutionary}.
In \FIG\ref{fig:evolutionary}(a), we set $m=3$ and prepare
10 memory-one players of each of the $4^4=256$ types
and 25 BM players in the beginning. Therefore, $N=2560+25=2585$.
In \FIG\ref{fig:evolutionary}(b), we set $m=5$ and prepare
two memory-one players of each of the $6^6=1296$ types and
25 BM players in the beginning. Therefore, $N=2592+25=2617$.
In both cases, the BM players can invade the population of memory-one
players to eventually become dominant. Within the memory-one players,
those with large $p_{\rm CC}$ tend to survive at early stages of
evolutionary dynamics before they are overwhelmed by the BM player.

\section{Discussion}

We numerically analyzed the behavior of a BM model in the iterated PD.
Our model is a modification of the BM model used by Macy and Flache (2002) 
such that the probability of cooperation in the
next round is made sensitive to the reinforcement signal obtained in
the current round. Our model is also a close variant of the models used by
Karandikar \textit{et al.} (1998) and
Posch \textit{et al.} (1999).
When the adaptation of the aspiration level
is not too fast, the modified BM player mutually cooperates
with a large probability. The BM player also performs efficiently against
reactive strategies and in evolutionary dynamics in a population
comprising various memory-one strategies.
Up to our numerical efforts, the results are 
robust against the error in the action implementation and the change in the payoff matrix
describing the PD.

The BM player performs at least comparably to
memory-one players such as GTFT and Pavlov,
which are strong competitors in the iterated PD.
Although the BM player is inferior to these strategies when playing against
TFT, it performs better 
than GTFT
against other strategies including ALLC and ALLD.
In an evolutionary context, naively cooperating with
ALLC allows it to prosper by a neutral drift,
which eventually invites the invasion of
malicious players such as ALLD. Therefore, it is important to be able
to exploit ALLC for a strategy to survive in evolutionary dynamics
\cite{Nowak1993Nature}. This property is not satisfied by
TFT \cite{Axelrod1984book},
GTFT \cite{Nowak1992Nature_gtft}, Pavlov \cite{Kraines1989,Nowak1993Nature},
and BM model with a fixed aspiration level
\cite{Macy1991AJS,Macy1996SMR,Posch1999RoyalB,Macy2002pnas}. In contrast, 
our BM player as well as
the temporal difference learner \cite{MasudaOhtsuki2009BMB} are capable of exploiting ALLC.

As a model of humans and other animals in iterated games,
a learning strategy may be generally disadvantageous
as compared to simpler learning strategies and non-learning strategies in at least two aspects.
First, the number of rounds before learning is established may be large.
For example, the temporal difference learning, a type of
reinforcement learning, must be implemented with
a very small learning rate to realize
mutual cooperation \cite{MasudaOhtsuki2009BMB}. In contrast,
the learning of the modified BM model is completed in
one to some hundreds of rounds. This speed of learning is comparable to
that of other learning models in which mutual cooperation
is obtained within ten to hundred rounds
\cite{Macy1991AJS,Macy1996SMR,Erev1998AER,Erev2001toolbox,Hauert2002jtb_learn,Macy2002pnas}. 
Second, humans or other animals subjected to
social dilemma situations may not implement a complex learning strategy.
In this aspect, the BM model with the
adaptive aspiration level, both the original ones
\cite{Karandikar1998JET,Posch1999RoyalB,Macy2002pnas} and ours, has a clear advantage.
The BM model is simpler than many learning models including
the temporal difference learning
\cite{Sandholm1996BioS,MasudaOhtsuki2009BMB},
fictitious play \cite{Erev1998AER,Fudenberg1998book,Camerer2003book},
genetic algorithms
\cite{Macy1996SMR}, and artificial neural networks
\cite{Macy1996SMR,Sandholm1996BioS,Taiji1999PhysicaD,Gutnisky2004AL}.

The memory-one strategy, for example, can be regarded as a
reinforcement learning because the probability of cooperation is a function
of the outcome in 
the previous round. This is also the case for analogous strategies with
longer memory \cite{Lindgren1991ALife}. Nevertheless, in
this study, we are concerned with the cases where
the probabilities of cooperation conditioned by
the recent results of the game adapt over time.
In the case of the BM model, adaptation is realized by the dynamic
aspiration level.
Learning
players in this restricted sense cope with various types of opponents more
flexibly than the memory-one strategy or its extension with longer memory.
For example, we showed that the mean aspiration level of the 
BM player is almost equal to the mean payoff against different reactive
strategies (\SEC\ref{sub:reactive}). This result indicates that
the BM player flexibly behaves as different types of win-stay lose-shift
strategists depending on the opponent.
Learning in games is a recent outstanding issue involving interdisciplinary 
research fields such as 
behavioral game theory and neuroeconomics
\cite{Fudenberg1998book,Camerer2003book,Glimcher2009book}.
Because the effect of learning is evident in laboratory experiments
\cite{Camerer2003book,Glimcher2009book}, it may be important to consider
individual learning in addition to evolution to understand
the behavior of agents, particularly that of humans,
in social dilemma situations.
Our model, which is simple yet competitive in the PD,
may be used for examining various problems with regard to relationships
between learning and cooperation in social dilemma situations.

\section*{Acknowledgments}

We thank Shoma Tanabe for careful reading of the paper.
N.M. acknowledges the support from the Grants-in-Aid for Scientific
Research (No.\ 20760258).
M.N. acknowledges the support and the Grants-in-Aid for
Scientific Research from JSPS, Japan.

\section*{Appendix: Reactive strategy against itself}

When the focal player obeys the reactive strategy with parameters
$\overline{p}$ and $\overline{q}$,
the long-term behavior of the focal player against
the reactive strategy with parameters $p$ and $q$ can be analytically calculated
\cite{Nowak1989AMC,Nowak1990Acta,Nowak1990TPB,Nowak2006book}.
The probability that the focal
player cooperates is given by
\begin{equation}
s_1=\frac{q(\overline{p}-\overline{q})+\overline{q}}
{1-(\overline{p}-\overline{q})(p-q)}.
\label{eq:s_1}
\end{equation}
The mean payoff of the focal player is given by
\begin{equation}
Rs_1s_2 + Ss_1(1-s_2)+T(1-s_1)s_2+P(1-s_1)(1-s_2),
\end{equation}
where $s_1$ is given by \EQ\eqref{eq:s_1} and
\begin{equation}
s_2=\frac{\overline{q}(p-q)+q}
{1-(\overline{p}-\overline{q})(p-q)}.
\label{eq:s_2}
\end{equation}

\newpage
\clearpage

\begin{figure}
\begin{center}
\includegraphics[height=12cm]{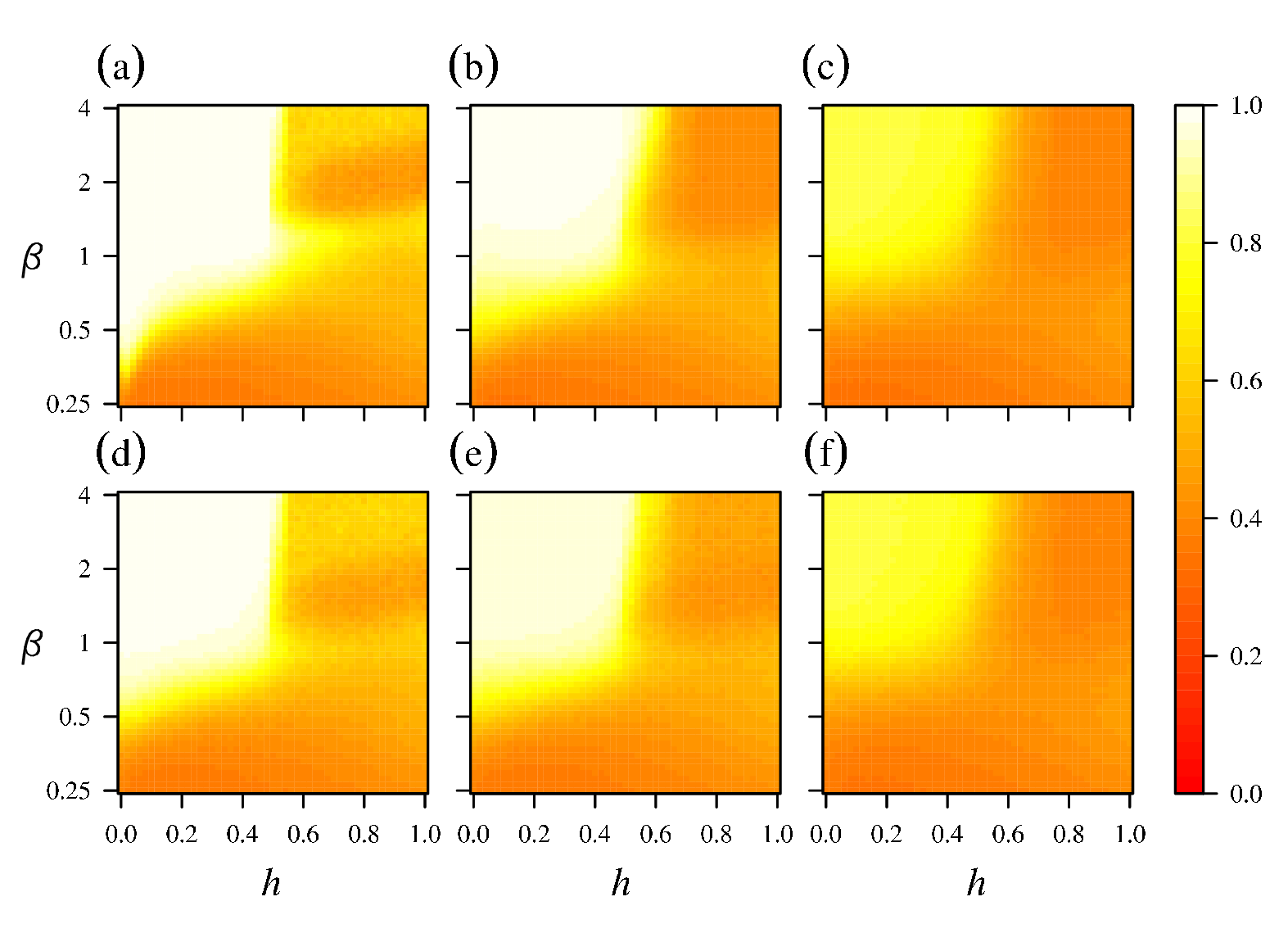}
\caption{\setlength{\baselineskip}{0.77cm}
Fraction of cooperation of the BM player playing against another BM
player. The number of rounds is equal to
(a--c) $t_{\max}=1000$ and (d--f) $t_{\max}=100$. 
The probability of the misimplementation of the action 
is equal to (a, d) $\epsilon=0$, (b, e) $\epsilon=0.01$, and (c, f) 
$\epsilon=0.1$. We set $R=3$, $T=5$, $S=0$, and $P=1$, and vary $h$ and $\beta$.}
\label{fig:BM-BM}
\end{center}
\end{figure}

\clearpage

\clearpage

\clearpage

\clearpage

\clearpage

\clearpage

\end{document}